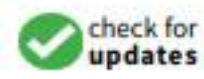

# Automated detection of circadian-dependent epileptic biomarkers for seizure localization using machine learning and signal processing

**Mehdi Zekriyapanah Gashti**[1+]
**Mostafa Mohammadpour**[2]
**Hassan Eshkiki**[3]
**Vahid Ghanbarizadeh**[4]

[1]*Department of Data Science and Business Intelligence, MZG Engineering Consultancy, Hamburg, Germany.*
[1]*Email: gashti@ieee.org*
[2]*Department of Computational Perception, Johannes Kepler University, Linz, Austria.*
[2]*Email: m.mohammadpour@ieee.org*
[3]*Department of Computer Science, Swansea University, Swansea, United Kingdom.*
[3]*Email: h.g.eshkiki@swansea.ac.uk*
[4]*Department of Computer Engineering, Florida Atlantic University, Boca Raton, United States.*
[4]*Email: vghanbarizad2023@fau.edu*

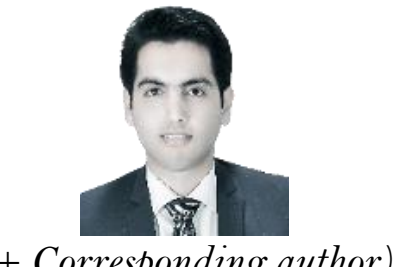

*(+ Corresponding author)*

## ABSTRACT

Accurate localization of the seizure onset zone (SOZ) is essential for successful epilepsy surgery, yet the reliability of commonly used interictal biomarkers is limited by temporal variability and behavioral state. This study aims to investigate the circadian and sleep-dependent dynamics of epileptic biomarkers and to identify conditions that maximize seizure localization precision. Long-term intracranial EEG recordings from nine patients with drug-resistant focal epilepsy were retrospectively analyzed using automated signal processing and machine learning techniques. Interictal spikes, spike sequences, high-frequency oscillations (HFOs), and pathological HFOs were automatically detected, while sleep and wake states were classified using the alpha–delta power ratio. Biomarker rates, spatial distributions, and localization accuracy were quantitatively evaluated using Euclidean distance relative to the clinically defined SOZ. The results show that all biomarkers exhibit significantly higher rates during sleep, with pronounced early-morning peaks. Importantly, spike sequences and pathological HFOs demonstrated superior spatial precision compared to conventional spikes or HFOs alone. Mean distances to the SOZ were substantially lower for pathological HFOs and spike sequences, with statistically significant differences among biomarkers (ANOVA, $p < 0.001$). These findings demonstrate that sleep-state analysis, particularly using propagated spike sequences and pathological HFOs, substantially improves SOZ localization accuracy. The proposed framework provides practical guidance for sleep-focused presurgical EEG analysis and supports the development of automated and clinically efficient seizure localization systems.

**Contribution/Originality:** This study contributes to existing literature by systematically analyzing circadian and sleep-dependent variability of epileptic biomarkers. It introduces a unified automated signal processing and machine learning framework. This study is among the few evaluating spike sequences and pathological HFOs jointly. The contribution shows HFOs and sequences improve seizure onset zone localization.

## 1. INTRODUCTION

Epilepsy is a neurological disorder affecting about 1–2% of the global population. Epileptic seizures are caused by abnormal neural discharges originating from the cerebral cortex. Many patients with epilepsy can be effectively managed with anti-epileptic medications. However, for those who do not respond to these medications, epilepsy surgery offers a potential path to seizure freedom. In these patients, the epileptogenic zone, crucial for surgical planning, is defined by the seizure onset zone (SOZ). This zone is considered the gold standard marker and is typically identified through presurgical evaluation of intracranial electroencephalography (iEEG) data [1].

Localizing the area responsible for seizure onset is challenging because prolonged intracranial recordings are needed to capture seizure onset, which usually takes several days in the epilepsy monitoring unit (EMU) [2]. Interictal epileptiform discharges (IEDs) and HFOs serve as promising biomarkers for localizing SOZ in epilepsy patients [3]. Unlike waiting for seizures to occur, these abnormal electrical patterns provide valuable insights into the irritative zone and areas of increased neuronal excitability, potentially expediting the identification of the SOZ during the presurgical evaluation process [4].

Interictal spikes are short electrographic discharges observed in patients with epilepsy during periods between seizures, serving as a clinical biomarker for epilepsy [5]. Clinically, the identification of spike locations plays a pivotal role in mapping epileptic networks for surgical planning [6, 7]. Notably, surgical outcomes tend to improve when regions characterized by frequent interictal spikes are targeted for resection [8-10]. This approach aligns with the strategy of resecting areas indicating abnormal electrographic activity, aiming to alleviate the impact of epilepsy and enhance the patient's overall prognosis [11].

High-frequency oscillations (HFOs) serve as biomarkers for detecting epileptogenic tissue and play a critical role in cortical function, particularly within the motor and sensory cortex. HFOs are classified into three categories: high-gamma (65-100 Hz), ripple (80-250 Hz), and fast ripple (250-500 Hz) [12, 13]. These oscillations are associated with both task-induced activities and the generation of seizures, characterized by notable spatial and temporal resolution. Typically, high-gamma and ripple oscillations are observed in the neocortex, whereas fast ripples are predominantly found in the hippocampus [3]. HFOs can also be recorded during the interictal state of an epileptic seizure, during rest, or in response to specific tasks. To be clearly distinguishable from background noise, HFOs must have at least six peaks that exceed three standard deviations from the mean [13]. Despite their clinical significance, the precise characteristics and definitions of clinically relevant HFOs remain poorly defined in the medical literature. The relationship between the removal of areas generating HFOs and positive surgical outcomes suggests the potential utility of HFOs as a marker for epileptogenicity [14-16], and it has been shown that their rate tends to be higher in SOZ areas [16].

The HFOs are recognized to exist in physiological or normal HFOs (nHFO) and pathological HFOs (pHFO) [17]. HFOs linked with IEDs are deemed more pathological due to the abnormal nature of IEDs, which are not typical physiological events [18]. Conversely, spontaneously occurring HFOs are generally viewed as physiological, reflecting normal physiological processes in the absence of IEDs [19]. The potential of classifying normal HFOs and pathological HFOs in cases of focal epilepsy has been demonstrated [20, 21]. They also revealed the correlation of pHFO with the SOZ and surgical outcomes. Spontaneous physiological ripples in the human visual cortex were identified [22]. These ripples are frequently observed in eloquent cortex regions. Differences in the relationship between slow waves and HFOs in normal and epileptic brain regions were observed, highlighting their distinct origins [22]. This finding holds practical importance as it enhances the differentiation between electrodes recording from normal and epileptic brain areas.

Prolonged intracranial recordings often yield a substantial amount of spike data. Studies analyzing recordings, typically spanning one or several nights, have consistently demonstrated a robust circadian pattern across all subjects, showing uniform peaks during normal sleep hours irrespective of the seizure-onset zone location [23, 24].

In this study, we explored the spatial dynamics of epilepsy biomarkers using intracranial EEG data to identify the optimal time periods for precisely localizing the seizure onset. To achieve this, we employed automated epilepsy biomarker detection and sleep-wake classification on a long-term iEEG dataset from patients with drug-resistant epilepsy. This approach allowed us to examine how sleep influences epilepsy biomarkers and determine the best times for data analysis to enhance SOZ localization accuracy. To support this objective, we conducted key experiments including validating the automated biomarker and sleep–wake detectors, analyzing biomarker spatial patterns across different sleep stages, and evaluating how temporal windows influence SOZ localization performance.

To address these limitations, this study investigates the circadian and sleep-dependent behavior of multiple epileptic biomarkers and evaluates their relative effectiveness for seizure onset zone localization. Specifically, we conduct three key experiments: (i) automated detection of interictal spikes, spike sequences, HFOs, and pathological HFOs from long-term intracranial EEG recordings; (ii) classification of sleep and wake states using alpha/delta ratio features; and (iii) quantitative comparison of biomarker localization accuracy using Euclidean distance to the clinically defined SOZ. This integrated experimental design enables a systematic assessment of how temporal state and biomarker type jointly influence localization performance.

The main contributions of this work are summarized as follows. First, we present an automated framework that jointly analyzes circadian dynamics, sleep state, and multiple epileptic biomarkers from long-term intracranial EEG recordings. Second, we demonstrate that propagated spike sequences and pathological HFOs provide significantly higher spatial precision for seizure onset zone localization than conventional spike or HFO biomarkers, as quantified using distance-based metrics. Third, we show that sleep, particularly early-morning data, offers the most informative temporal window for SOZ localization. These findings provide practical guidance for sleep-focused presurgical EEG analysis and lay the groundwork for future real-time and automated seizure localization systems.

The remainder of this paper is organized as follows. Section 2 describes the patient cohort, data preprocessing steps, and the automated methods for biomarker detection and sleep classification. Section 3 presents the experimental results, including circadian analysis, sleep-dependent biomarker variations, and quantitative localization performance relative to the seizure onset zone. Finally, Section 4 concludes the paper and outlines directions for future research.

## 2. METHOD

### *2.1. Patient Selection and Data Preparation*

We selected nine patients with drug-resistant epilepsy who underwent intracranial EEG recordings as part of their presurgical evaluation at the University of Pennsylvania (ieeg.org). All patients included in the study had at least 24 hours of iEEG data, experienced at least one seizure, and had sleep-wake annotations available. Data from each patient were segmented into 10-minute blocks, and seizure segments were removed from the analysis. Channels with significant deviations from baseline (median) voltage, high variance, excessive 60 Hz noise, or unusually high standard deviation compared to other channels were identified and excluded as bad channels. Subsequently, a common average reference (CAR) source derivation was applied to remove spatially coherent noise across electrodes.

### *2.2. Epilepsy Biomarker Detection*

An established method outlined in Janca et al. [25] was utilized to detect interictal spikes. To ensure uniform filter characteristics, signals with higher sampling rates were first down-sampled to 200 Hz. Each channel was then filtered using zero-phase filtering within the 10-60 Hz range, applying a combination of 8th-order type II Chebyshev high-pass and low-pass digital filters. Subsequently, the instantaneous envelope of each filtered channel was computed by taking the absolute value of the Hilbert transform. Spikes typically induce an energy increase, manifesting as a peak in the envelope within the 10-60 Hz frequency band. The statistical distribution of the signal envelope was calculated for each epoch, and a statistical model was fitted using maximum likelihood estimation (MLE). Consequently, the mode and median of the normalized (log-normal distribution) data were utilized to establish a threshold for distinguishing epochs containing spikes from those exhibiting background activity. Figure 1 provides an example of the detected spikes and HFOs in the iEEG recordings. Spikes are shown after applying a 0.5 Hz high-pass filter to highlight sharp transient events, while HFOs are displayed using an 80–250 Hz band-pass filter to isolate the fast oscillatory activity characteristic of these biomarkers. Figure 1-A suggests spikes in an iEEG trace.

A previously published method [26], the Hilbert Detector (HIL), was employed to detect HFOs [27]. In this approach, the signal undergoes initial filtering within the 80-250 Hz in the ripple band HFOs, followed by the computation of the envelope using the Hilbert transform. To be regarded as a valid HFO, an event must satisfy two conditions: firstly, the local maximum for each event must surpass a threshold of 5 standard deviations (SDs) of the envelope, originally calculated over the entire recording or from a specific time interval. Secondly, each identified HFO must have a minimum duration of 10 milliseconds. Figure 1-B suggests iEEG signal traces for HFOs in the ripple band.

A recent study by Mohammadpour et al. [28] proposed an unsupervised framework for detecting high-frequency oscillations (HFOs) in the ripple and fast ripple bands by combining S-transform-based time-frequency analysis with comprehensive multi-domain feature extraction and clustering. This method leverages statistical, spectral, and image-based features to effectively differentiate true HFOs from artifacts and background activity, without relying on supervised learning or manual annotation. The approach was rigorously validated on both controlled simulation data and clinical datasets from epilepsy patients, demonstrating high detection accuracy and a strong correlation between HFO resection and favorable post-surgical outcomes. These findings further substantiate the role of fast ripples as robust biomarkers of epileptogenic zones and support the clinical relevance of automated HFO detection in epilepsy surgery planning.

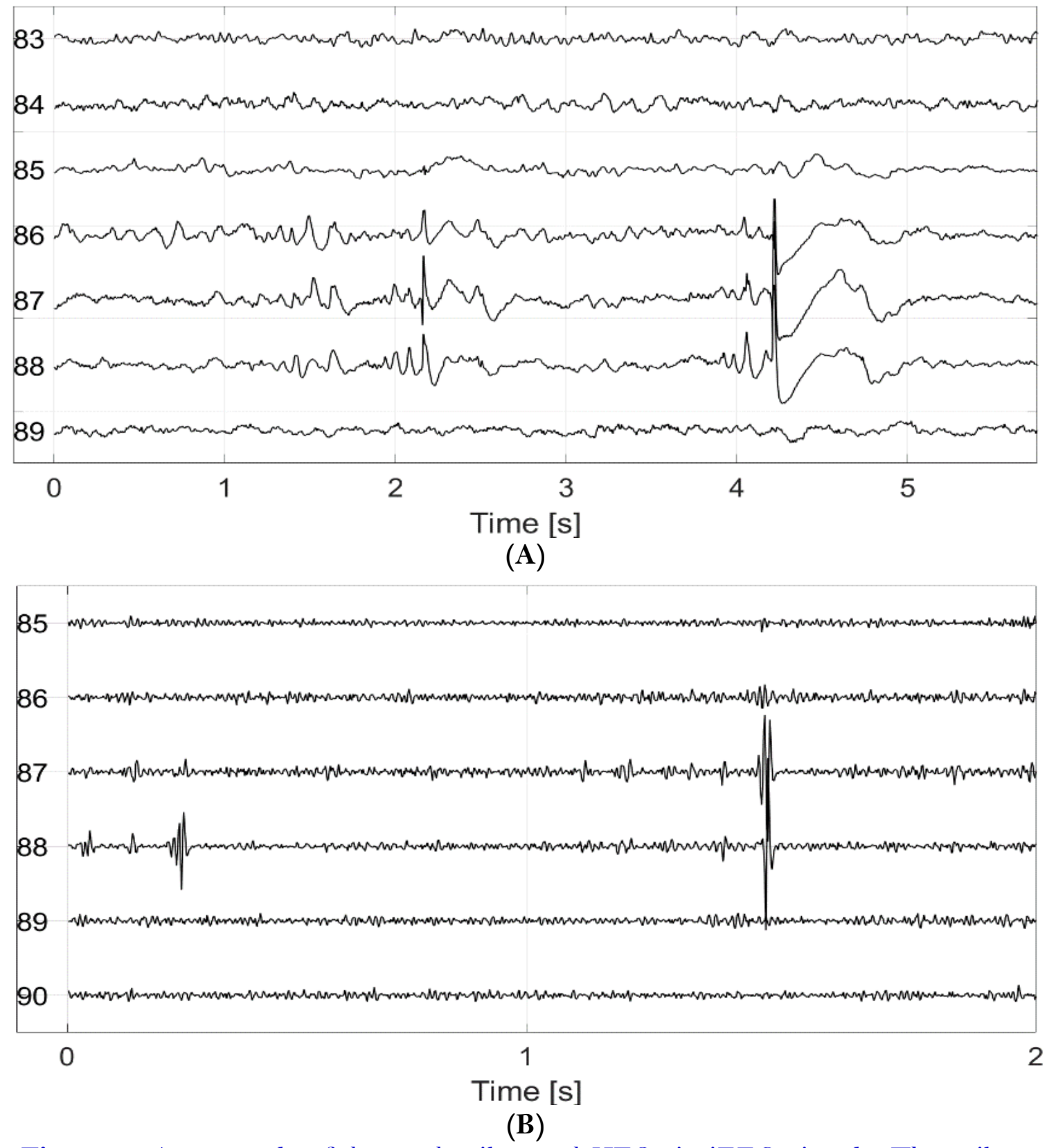

**Figure 1.** An example of detected spikes and HFOs in iEEG signals. The spike traces are filtered with a high-pass filter at 0.5 Hz, while HFO signals are band-pass filtered within the range of 80-250 Hz.

After detecting individual events (Spikes and HFOs), the subsequent step involved identifying the event sequences to investigate their propagation characteristics, particularly when interictal spikes occurred simultaneously across multiple electrodes in close temporal proximity. A previously described algorithm was employed to achieve this [29]. In brief, the initial spike within a potential sequence was designated as the leader. Subsequent spikes occurring within 50ms of the leader or 15ms of the preceding spike were added to the sequence. Sequences comprising fewer than five spikes were excluded. Moreover, sequences were discarded if 55% of the spikes within the sequence occurred within 2ms of each other, as these were more likely to be considered artifacts upon visual inspection. Figure 2 shows iEEG signal traces for spike sequences. Pathological HFOs can be induced to occur concurrently, superimposed on interictal discharges [30]. Since interictal spikes are inherently pathological, the presence of additional events, such as HFOs within them, can be termed pathological HFOs (pHFO). Figure 3 shows traces of pathological and physiological HFOs.

### *2.3. Sleep Analysis*

For sleep classification, 1-minute segments of labeled sleep and wake states were selected for each patient. The alpha-delta ratio (ADR) feature, defined as the ratio of power in frequency bands alpha (8-13 Hz) and delta (1-4 Hz), was calculated for each segment. Calculating the average value of the ADR in all channels and normalizing them over the overall time segments yielded a single ADR value for each segment and patient. Subsequently, using a threshold-based binary classification, segments were classified as representing either sleep or wakefulness by determining a potential threshold based on the ADR. Subsequently, the circadian rhythm of epilepsy biomarkers was examined to determine its impact on these biomarkers. To achieve this, the occurrence of these biomarkers was counted in 144 ten-minute intervals throughout the day (six intervals per hour over 24 hours) for each patient. To enhance classification robustness, machine learning techniques combining multiple classifiers with optimization algorithms have been proposed. These validated methods informed the design and implementation of the classification approach used.

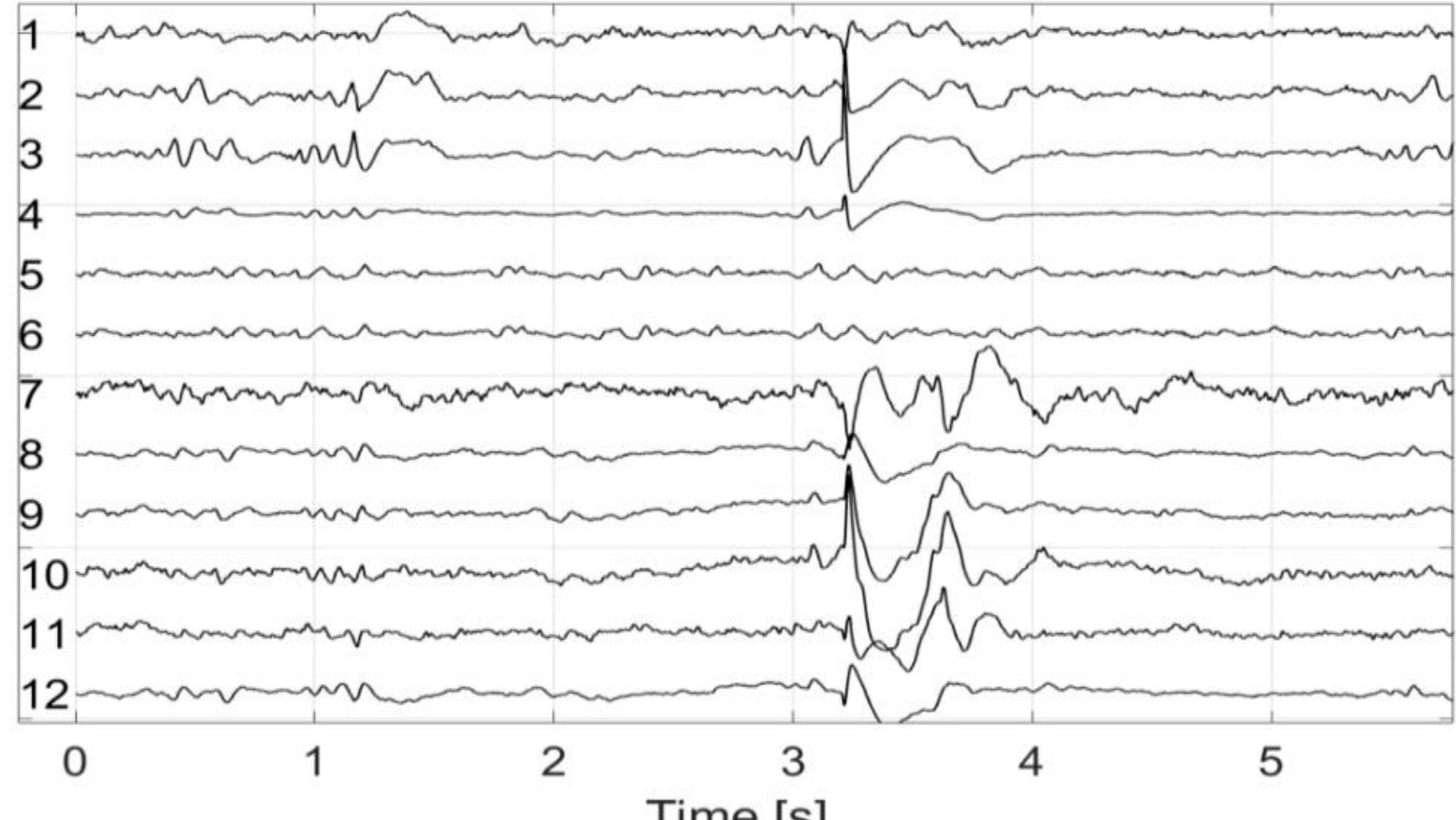

**Figure 2.** An example of a spike sequence in an iEEG signal that was high-pass filtered at 0.5 Hz.

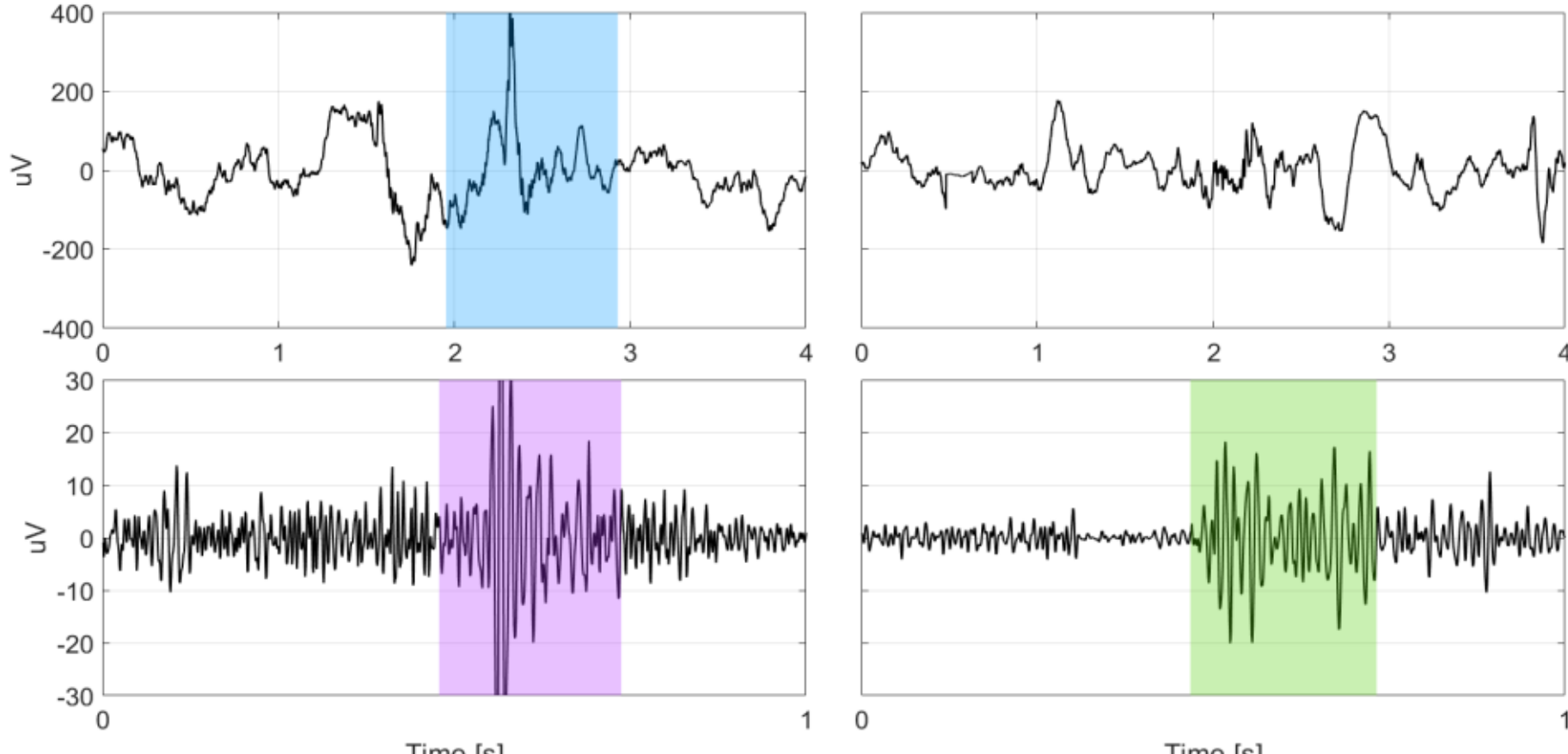

**Figure 3.** Example of pathological and physiological HFOs: The upper traces display iEEG data filtered with a high-pass filter at 0.5 Hz, while the lower traces show band-pass filtering in the range of 80-250 Hz. In the left traces, a spike coincides with HFOs, indicating pathological activity. In contrast, the right trace shows isolated HFOs, where there is no spike, which is considered physiological or normal.

## 3. RESULTS AND DISCUSSION

### *3.1. Clinical Characteristics and Epileptic Biomarkers Detection*

A total of 216 hours (9 days) of iEEG data were studied across nine patients, involving 808 electrodes in total. During analysis, 7922 spikes and 2803 HFOs were detected (events/channel no./min). Subsequently, the application of a sequence detector for spike events identified 6267 spike sequences (events/channel no./min). An overlap analysis between spike and ripple HFOs to identify pHFO events revealed 1343 pHFO (events/channel no./min).

### *3.2. Sleep Classification*

The receiver operating characteristic (ROC) curve and the area under the ROC curve (AUC) are used to evaluate the performance of binary classification models. When comparing the classification output of sleep and wake with the annotated class of iEEG data, the AUC was 0.84 for sleep classification. Figure 4 suggests an ROC curve illustrating the accuracy of ADR (AUC = 0.84) in classifying sleep/wake states for nine patients with annotated sleep/wake data. Figure 5 depicts a polar histogram illustrating the amount of iEEG segments classified as sleep state across different times of the day. The peak sleep periods were observed in the early morning, specifically between 3 and 7 am.

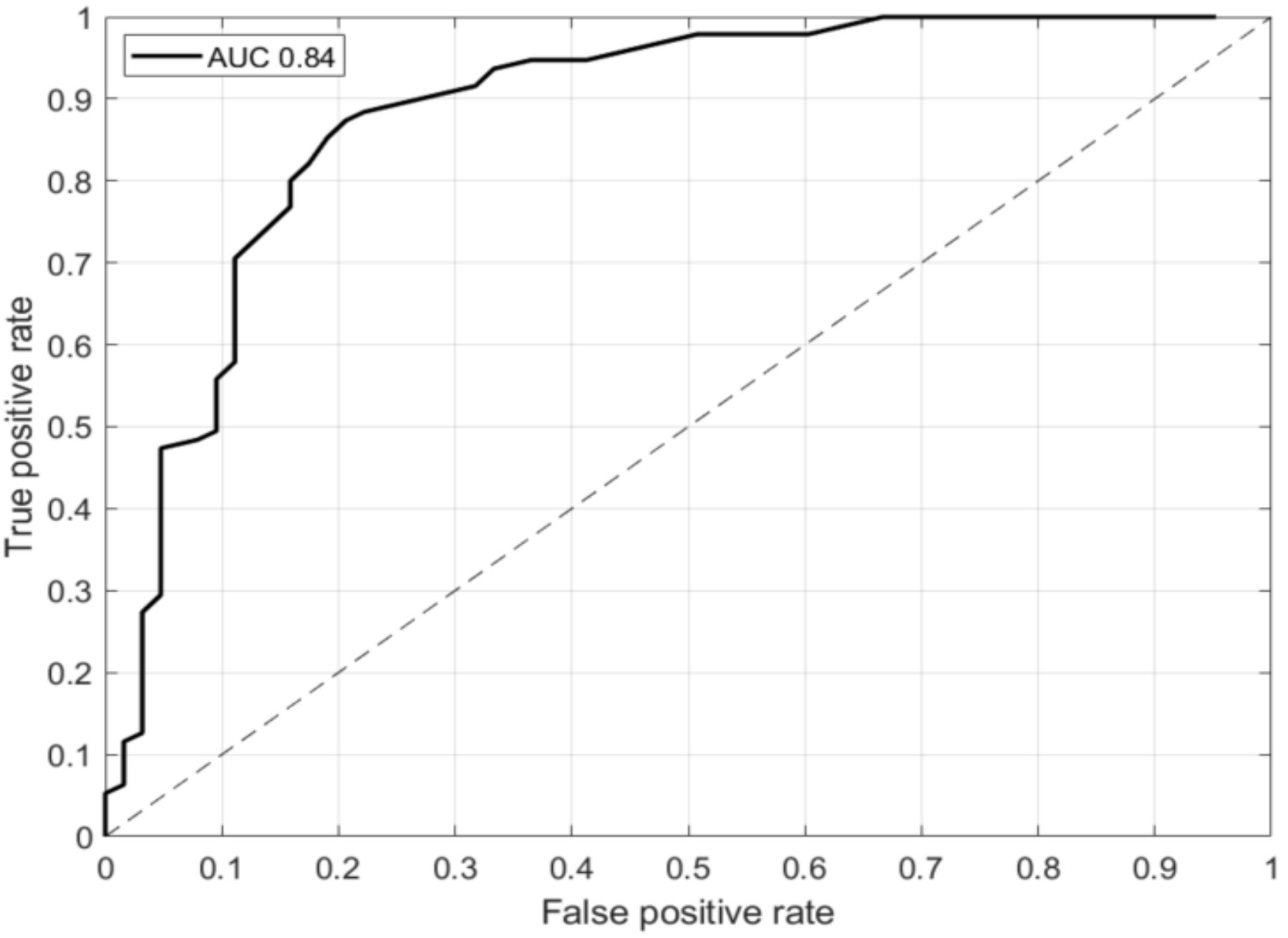

**Figure 4.** A ROC curve suggests the accuracy of ADR in predicting sleep or wake states for nine patients.

### *3.3. Variations of Biomarkers with Sleep*

The analysis of the iEEG data over a 24-hour period revealed a circadian rhythm peak for the biomarkers. The finding supports the idea that epilepsy biomarker rates follow a distinct circadian pattern. Examination of the rates of the four biomarkers spike, spike sequence, HFO, and pHFO by time of day revealed non-uniform distributions for each. The Rayleigh test analysis of patients' circular means showed significant non-uniformity for all biomarkers.

Specifically, spike sequence and pHFO rates exhibited the highest non-uniformity ($z=143.5$, $p<0.001$ and $z=143.8$, $p<0.001$, respectively). Additionally, HFO and spike rates also showed a significant non-uniform distribution ($z=142.2$, $p<0.001$ and $z=141.9$, $p<0.001$, respectively).

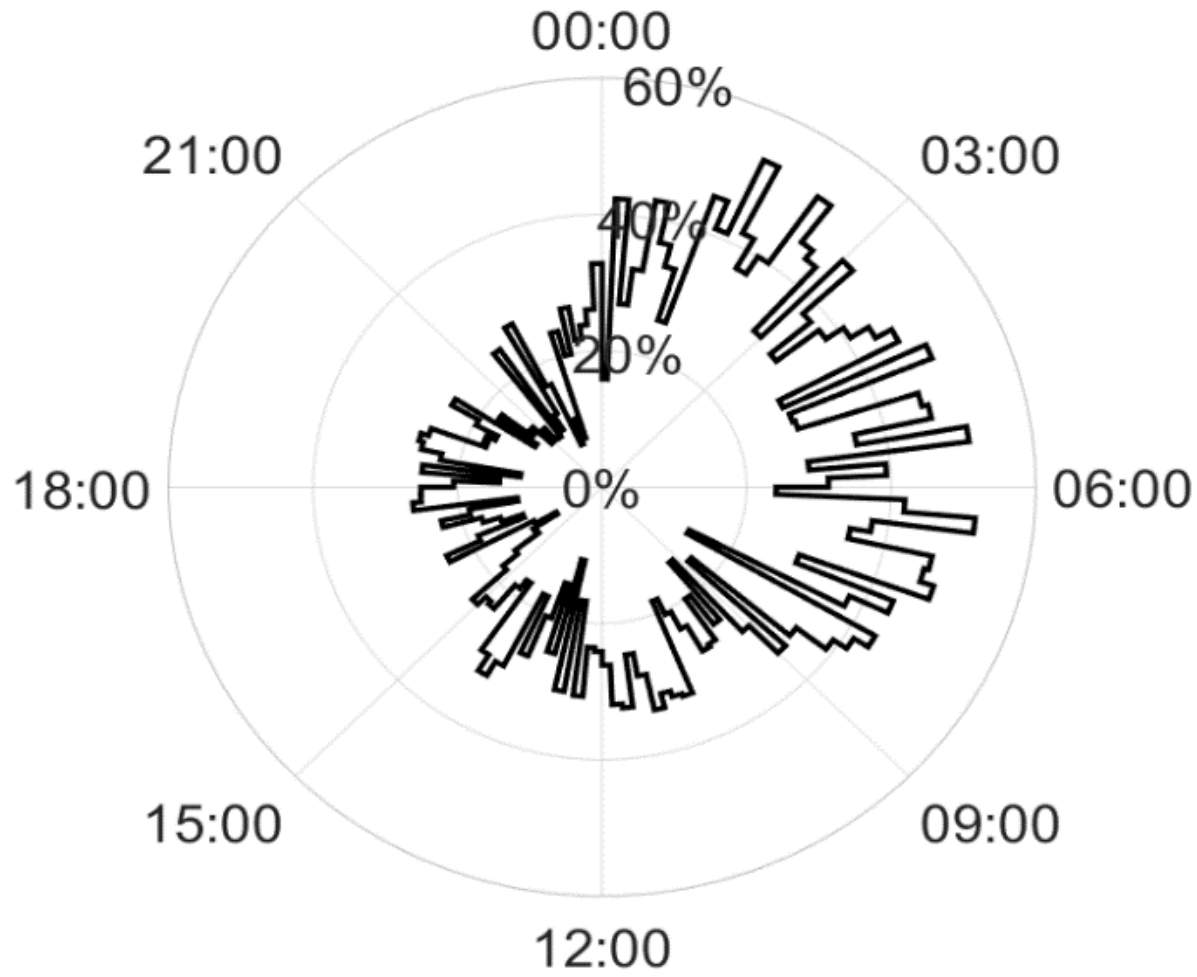

**Figure 5.** A polar histogram shows the amount of patient EEG segments predicted to be in a sleep state regarding the time of day, with peak sleep times occurring at the beginning of the day.

Visually, spike, spike sequence, and pHFO rates seem to be highest in the early morning and reach their lowest points in the late morning, afternoon, and evening (Figure 6). Biomarker rates from sleep and wake states are illustrated in Figure 7. Significantly higher spike rates were observed during sleep (19.3±10.66 events per minute) compared to wakefulness (9.6±6.3 events per minute) (Wilcoxon signed-rank test: $p < 0.005$). HFO rates showed nearly equal occurrences during sleep (14.4±10.8 events per minute) and wakefulness (14.5±12.1 events per minute) (Wilcoxon signed-rank test: $p < 0.5$). Spike sequence rates were higher during sleep (13.3±7.9 events per minute) than wakefulness (8.8±6.6 events per minute) (Wilcoxon signed-rank test: $p < 0.005$). Similarly, pHFO rates were higher during sleep (5.2±4.4 events per minute) compared to wakefulness (2.1±2.0 events per minute) (Wilcoxon signed-rank test: $p < 0.005$).

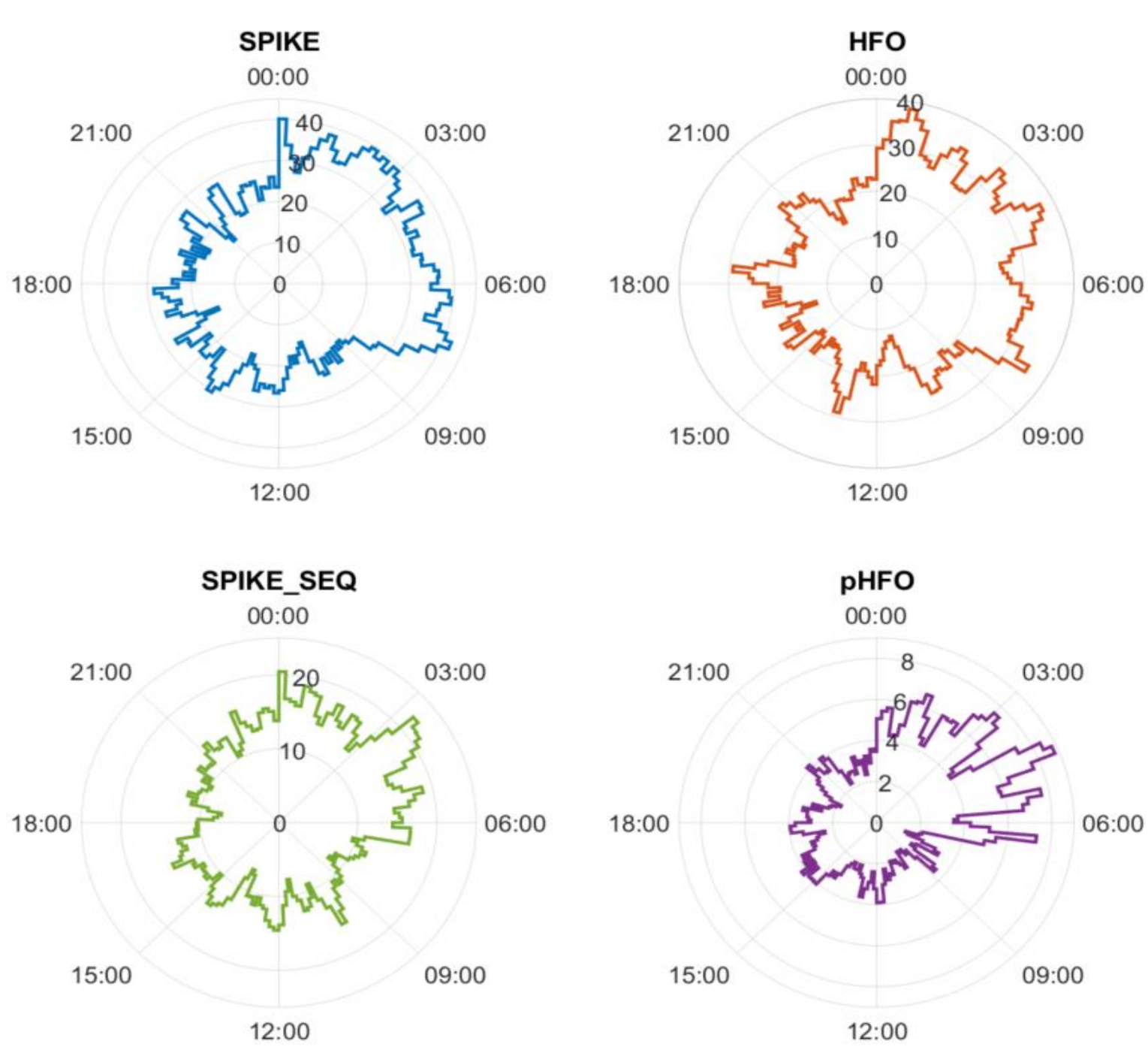

**Figure 6.** A polar histogram demonstrating the biomarker rates. Visually, the rates of all biomarkers seem to peak in the early morning and drop to their lowest levels in the late morning, afternoon, and evening. The pHFO biomarkers have the lowest rate but have sharper peaks early in the morning.

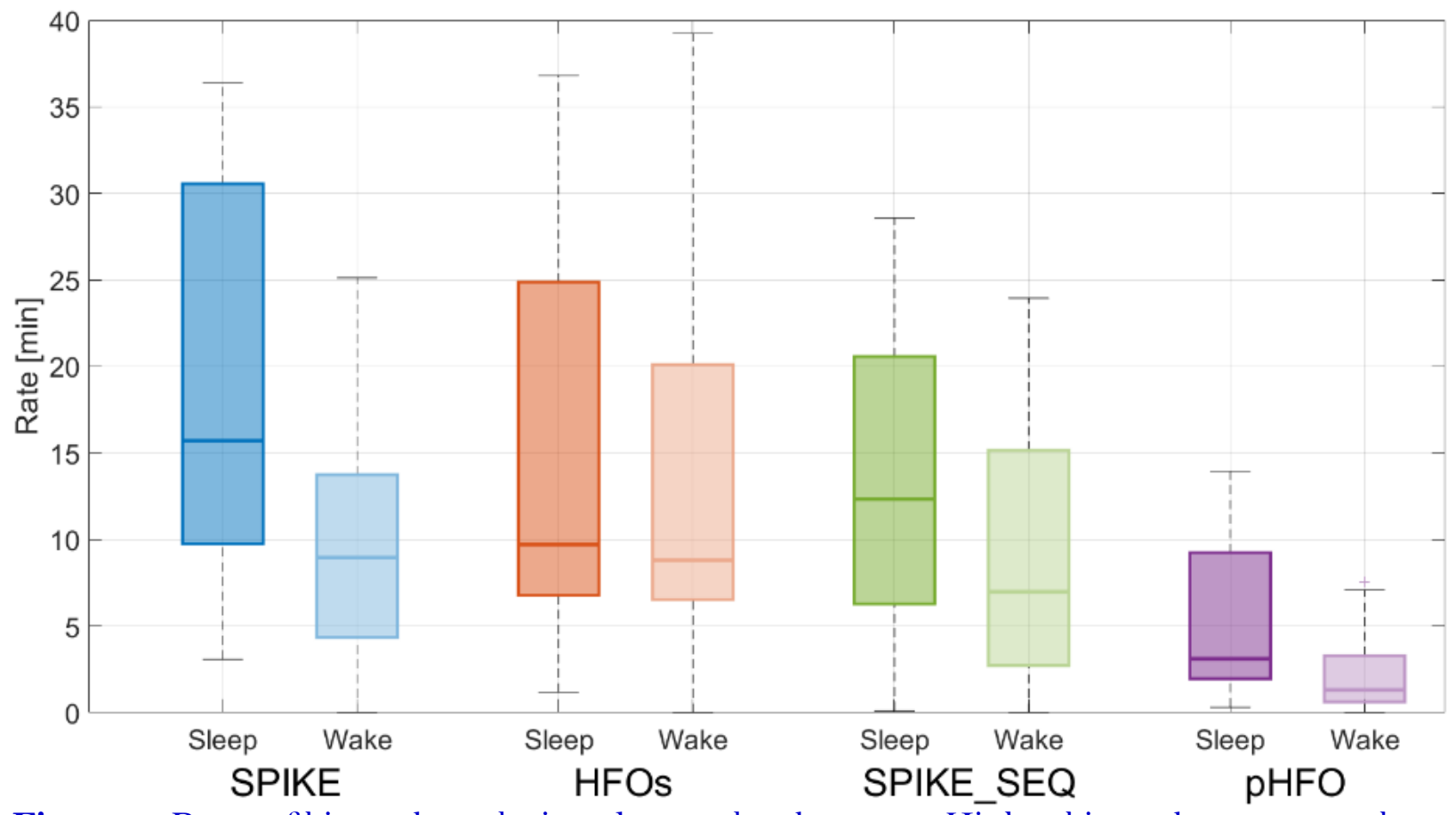

**Figure 7.** Rates of biomarkers during sleep and wake states. Higher biomarker rates can be seen during sleep compared to wakefulness.

### *3.4. Relationship between Biomarkers' Rate and SOZ*

For all patients, one hour of sleep data starting from midnight was selected, and channels within the SOZ were compared with channels outside the SOZ. All channels within the SOZ include event spikes, HFOs, spike sequences, and pHFOs, whereas channels outside the SOZ mainly include spikes and ripples. Figure 8 suggests the biomarker rate in SOZ and non-SOZ. The mean rate of spike events in SOZ and non-SOZ channels was 16.2±5.3 and 11.2±4.0 per minute, respectively, with the mean rate significantly higher in SOZ channels ($F\ [1, 118]: 33.23$, $p<0.001$). Similarly, the mean rate of HFO events in SOZ and non-SOZ channels was 13.7±3.9 and 11.7±4.8 per minute, respectively, with a significantly higher rate in SOZ channels ($F\ [1, 118]: 6.47$, $p<0.001$). The mean rate of spike sequence events in SOZ and non-SOZ channels was 14.2±2.6 and 7.0±4.0 per minute, respectively, with a significantly higher rate in SOZ channels ($F\ [1, 118]: 131.87$, $p<0.001$). Lastly, the mean rate of pHFO events in SOZ and non-SOZ channels was 4.7±1.3 and 2.1±1.6 per minute, respectively, with a significantly higher rate in SOZ channels ($F\ [1, 118]: 88.89$, $p<0.001$).

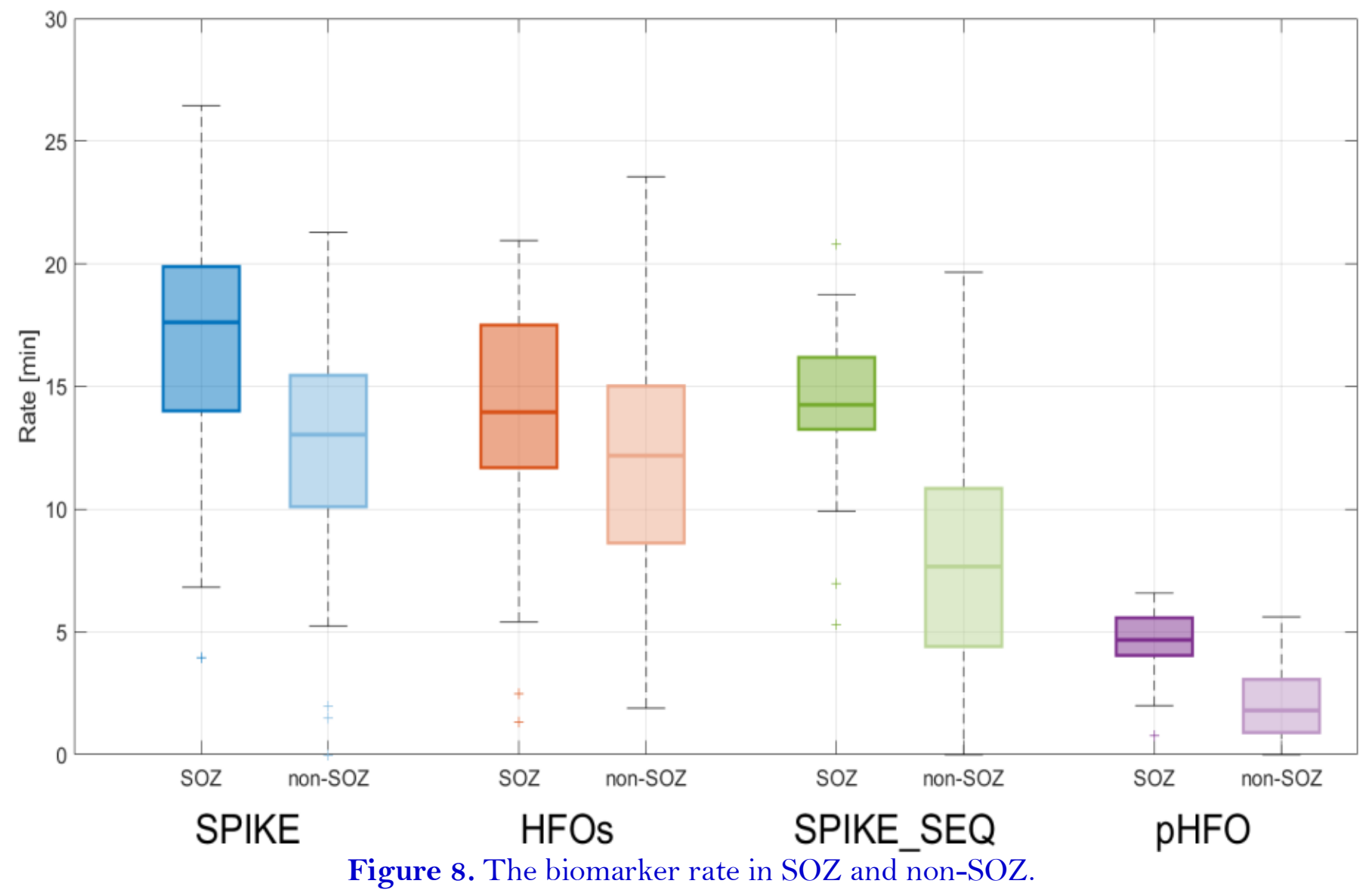

**Figure 8.** The biomarker rate in SOZ and non-SOZ.

### *3.5. Relationship between Biomarkers' Distance and SOZ*

In the earlier experiments, the percentage of detected events precisely located in seizure-generating channels was illustrated. However, a potential concern arises from considering neighboring channels to the seizure onset as non-seizure areas. Due to the neural propagation principle, interictal spikes are at times recorded at multiple sites with distinguishable latency, indicating rapid cortical propagation of spikes. The aim is to demonstrate that events may occur in the vicinity of the seizure-generating area. To achieve this, results comparing the distance of each individual channel displaying events to seizure onset were presented. For distance calculation, channels containing at least one event were selected, and the minimum distance from the chosen individual channel to channels indicating the seizure onset area was computed using Euclidean distance. Figure 9 illustrates the mean distance of all channels for each detected event for each individual patient to the SOZ. Even though the pHFO rate is low, we can see how close it is to the SOZ, demonstrating the effectiveness of using distance metrics to compare the epilepsy biomarkers network to the SOZ site. The mean distance results from Fig. 9 reveal that events obtained from spike sequence (19.3±16.6 mm) and pHFO (13.5±13.7 mm) are closer to seizure onset compared to independent events spike (24.5±20.9) and HFOs (22.6±19.7 mm). A one-way analysis of variance (ANOVA) was conducted to compare the distances of biomarkers to the SOZ. There was a statistically significant difference in the distances between these biomarkers, $F(3, 149482) = 1441.22$, $p < 0.001$. This indicates that the average distance to the SOZ significantly differs among the four biomarkers.

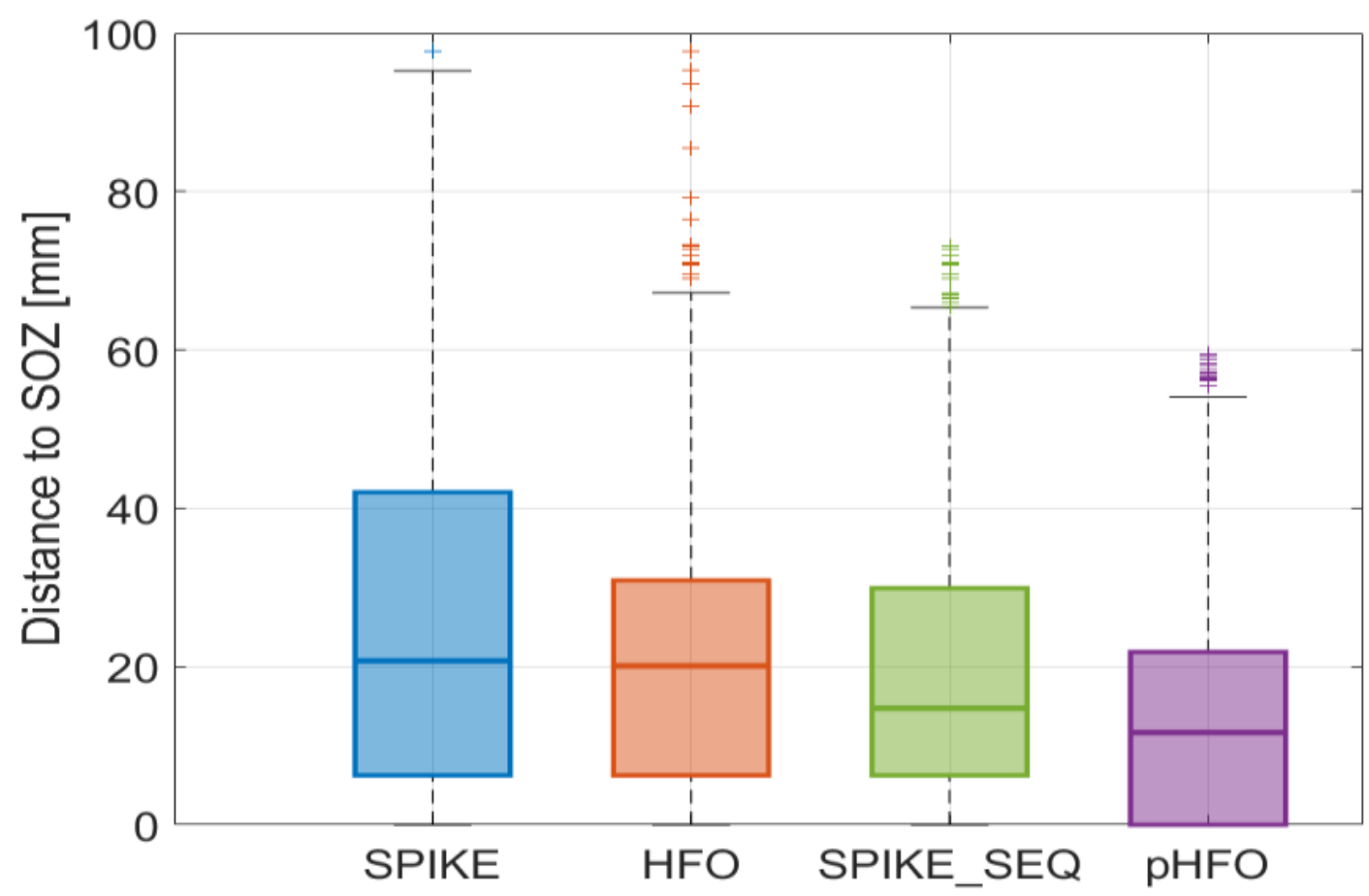

**Figure 9.** Distance of epilepsy biomarkers to seizure onset. The minimum distance from the individual channel with the biomarkers to the channels representing the seizure onset area was calculated using Euclidean distance.

Automated detection of epilepsy biomarkers and sleep/wake classification were applied to data from nine patients with focal epilepsy to determine how these biomarkers vary around sleep. We found that epilepsy biomarkers change with sleep, and spike sequences and pHFO, particularly in sleep, accurately identify seizure onset. Additionally, analyzing data in the early morning would be helpful for the detection of epilepsy biomarkers when their rates are highest. Experiments in this work indicate that events extracted through sequence analysis, which reveal their propagation patterns and pathological HFOs, could also serve as potential indicators for biomarkers of epileptogenesis. Furthermore, events identified through these two biomarkers were closer to the seizure location regarding electrode distance than those relying solely on spikes or HFOs.

Several studies have consistently demonstrated that circadian rhythm affects the epilepsy biomarkers [7]. This phenomenon was also observed in the patients of this study, especially spike sequence, and pHFO showed the best performance. However, spikes and HFOs were frequently observed in channels outside the seizure onset area, while pathological HFOs and spike sequences specifically occur in SOZ areas. Therefore, relying solely on spike and HFO would not have allowed for the accurate identification of the seizure onset.

In this study, to demonstrate the relationship between epilepsy biomarkers and seizure onset, we measured the distance of each individual event (which is in each electrode) from the seizure area using Euclidean distance. Similar to the previous comparison of event rate, we demonstrated that events were close to the seizure onset area for spike sequences and pathological HFOs, suggesting their potential role as biomarkers for seizure propagation.

In summary, the key findings of this study indicate that (1) epilepsy biomarkers exhibit clear sleep-dependent changes, (2) spike sequences and pathological HFOs provide more accurate and spatially specific indicators of seizure onset than spikes or HFOs alone, and (3) early-morning data contain the highest density of informative biomarkers and may therefore offer the best window for analysis.

While this study suggests the relationship between epilepsy biomarkers and sleep, it is important to acknowledge the limitations posed by the small sample size of only nine patients. This limited dataset may restrict the generalizability of our findings to the broader cohort with drug-resistant focal epilepsy. Future research should aim to include a broader population to validate these results and enhance their applicability. Additionally, multi-center studies could be beneficial, as they would increase the sample size and capture a more comprehensive range of demographic and clinical variability. Incorporating advanced statistical techniques and machine learning approaches could further enhance the analysis of larger datasets, potentially revealing more nuanced patterns in the relationship between sleep and epilepsy biomarkers. Future work could also integrate multimodal data (e.g., imaging or structural connectivity), investigate whether sleep-dependent biomarker dynamics can support real-time or adaptive localization tools, and apply advanced machine-learning methods to uncover more subtle patterns in biomarker propagation and circadian modulation. By addressing these limitations, subsequent studies could strengthen the evidence for the efficacy of spike sequences and pathological HFOs as reliable indicators of seizure onset.

## 4. CONCLUSION

Patients with drug-resistant focal epilepsy exhibit pronounced circadian and sleep-dependent fluctuations in epileptic biomarkers, which have direct implications for seizure onset zone (SOZ) localization. In this study, we systematically evaluated the spatial precision of multiple interictal biomarkers, including interictal spikes, spike sequences, high-frequency oscillations (HFOs), and pathological HFOs (pHFOs), across sleep and wake states using long-term intracranial EEG recordings. The results demonstrate that sleep, particularly early-morning periods, provides the most informative temporal window for SOZ localization. Although conventional spike and HFO rates increase during sleep, propagated spike sequences and pathological HFOs show substantially higher spatial precision when localization accuracy is quantified using distance-based metrics. These findings indicate that incorporating biomarker propagation characteristics and spike–HFO coupling leads to more reliable SOZ localization than relying on individual biomarkers alone. From a clinical perspective, these results strongly support targeted, sleep-focused analysis during presurgical intracranial EEG monitoring. Prioritizing sleep-state recordings, rather than uniformly analyzing extended monitoring periods, may streamline SOZ identification, reduce analysis time, and improve the efficiency of presurgical decision-making. This strategy has the potential to enhance surgical planning and contribute to improved postoperative seizure outcomes.

Future research should validate these findings in larger and more heterogeneous patient cohorts to assess their generalizability. Further investigation is also needed to clarify the neurophysiological mechanisms underlying the increased localization precision of pathological HFOs and spike sequences during sleep. Additionally, integrating multimodal data sources such as structural or functional neuroimaging and developing real-time or adaptive SOZ localization systems are important directions for translating these results into clinical practice.

**Funding:** The authors received no financial support for the research. The APC for this article was funded by "MZG Engineering Consultancy, Hamburg, Germany".
**Institutional Review Board Statement:** Not Applicable.
**Transparency:** The authors state that the manuscript is honest, truthful, and transparent, that no key aspects of the investigation have been omitted, and that any differences from the study as planned have been clarified. This study followed all writing ethics.
**Competing Interests:** The authors declare that they have no competing interests.
**Authors' Contributions:** All authors contributed equally to the conception and design of the study. All authors have read and agreed to the published version of the manuscript.